\RequirePackage[hyphens]{url}
\documentclass[%
 reprint,
 amsmath,amssymb,
 aps,
]{revtex4-2}

\usepackage{graphicx,amssymb,amsmath,hyperref}
\usepackage{xfrac}
\usepackage{courier}
\usepackage{booktabs}
\usepackage[hyphens]{url}
\usepackage{comment}
\usepackage{lineno}
\usepackage{cancel}
\usepackage[normalem]{ulem}
\usepackage{xcolor}
\usepackage{soul}

\begin{document}
 
\title{Beyond dynamic scaling: rare events break
universality}

\author{Ulysse Marquis$^{1,2}$}
\email{ulyssepierre.marquis@unitn.it}

\author{Riccardo Gallotti$^1$}

\author{Marc Barthelemy$^{3,4,5}$}
\email{marc.barthelemy@ipht.fr}

\affiliation{$^1$ Fondazione Bruno Kessler, Povo (TN), Italy}
\affiliation{$^2$ Department of Mathematics, University of Trento, Povo (TN), Italy}
\affiliation{$^3$ Universit\'e Paris-Saclay, CNRS, CEA, Institut de Physique Th\'eorique, 91191, Gif-sur-Yvette, France}
\affiliation{$^4$ Centre d’Analyse et de Math\'ematique Sociales (CNRS/EHESS) Paris, France}
\affiliation{$^5$ Complexity Science Hub, Vienna, Austria}

\begin{abstract}

Surface growth driven by non-monomeric deposition has remained largely unexplored. We investigate a model based on the deposition of blobs with a power-law size distribution $P(s)\sim s^{-\tau}$. We find that the critical exponents vary 
continuously with $\tau$, recovering Kardar--Parisi--Zhang behavior only for 
$\tau \ge 3$. For $\tau<3$, roughness scaling exhibits strong corrections and 
scale invariance breaks down. We show that this behavior originates from the 
emergence of a second dynamical length scale $\zeta$, corresponding to the 
linear size of the largest cluster, in addition to the usual correlation 
length $\xi$. The coexistence of these two relevant scales signals the 
breakdown of the usual Family--Vicsek scaling. These results point to a new 
phenomenology of surface growth beyond the standard scale-invariant paradigm.

\end{abstract}

\maketitle


\section{Introduction}


Surface growth is a central topic in non-equilibrium physics and has attracted sustained theoretical interest as a unifying framework to understand how interfaces roughen over time~\cite{Barabasi_Stanley_1995, kpz, Halpin_Healy_2015, Meakin1998}. In this context, the Kardar–Parisi–Zhang (KPZ) equation has emerged as a remarkably general description of interface dynamics: it is the most relevant stochastic evolution equation consistent with local growth normal to the interface, translational invariance, and rotational symmetry, and appears in a wide range of systems, from Burgers turbulence to directed polymers in random media. Experimentally, the surface growth lens was applied to a wide range of systems, including turbulent liquid crystals~\cite{takeuchi2010}, bacterial colonies~\cite{vicsek1990}, and tumor growth~\cite{bru1998}.

Traditionally, within the kinetic roughening paradigm, growth is modeled as the aggregation of small particles or as the smooth progression of an interface, described by stochastic evolution equations~\cite{odor2004} such as the Edwards--Wilkinson~\cite{Barabasi_Stanley_1995} (EW) or the KPZ~\cite{kpz} equation with white, uncorrelated noise. However, in many physical situations, growth proceeds through the attachment of extended clusters, leading to non-trivial amplitude fluctuations and strong spatio-temporal correlations. Examples include aerosol deposition~\cite{Friedlander2000}, liquid invasion in porous media~\cite{nolle1993, buldyrev1992}, and the expansion of cities~\cite{marquis2025}.

A natural question, then, is how such non-monomeric, laterally correlated growth mechanisms manifest. Considerable attention has been devoted to systems displaying anomalous scaling behavior: experimental studies of flows in porous media~\cite{rubio1989, horvath1991b}, slow paper combustion~\cite{Maunuksela1997}, bacterial colonies~\cite{vicsek1990}, oil-air interfaces~\cite{Soriano_2002,sardari2024}, electro-chemical deposition~\cite{galathara1992}, chemical vapor deposition~\cite{bowie2004, Son_2009} and wetting of paper~\cite{buldyrev1992} revealed values of the roughness exponent larger than expected in the EW or KPZ cases. Such observations indicate that fluctuations in real growth processes can be non-Gaussian, sometimes heavy-tailed, and are possibly accompanied by strong lateral correlations. These experimental findings have motivated theoretical and computational efforts to understand growth beyond the monomeric paradigm. Zhang~\cite{zhang1990, zhang1990_2} proposed a model with power-law distributed noise. This idea has been subsequently investigated in a discrete model in~\cite{buldyrev1991} and in the continuum in~\cite{lam1992, lam1993}. Krug~\cite{krug1991} predicted a simple relation between the exponent ruling the noise in the KPZ equation and the roughness exponent. It is however unclear how this maps to the discrete models, and whether this prediction is valid in small dimensions, which constitutes the relevant physical situation. These findings motivated the measure of the amplitude of the noise in fluid-fluid interfaces, which were found to be well described by a power-law with exponent~$\mu=2.7$~\cite{horvath1991a}. Further work investigated models with sphere deposition~\cite{csahok1992} and kinetic roughening of Hele-Shaw cells~\cite{Soriano_2002, sardari2024, Mirmahalle2025}. Medina et al.~\cite{medina1989} proposed a renormalisation-group analysis for the KPZ equation with temporally correlated noise and derived the values of the dynamic and roughness exponents. Chow~\cite{chow1997} proposed an explanation for anomalous values of the roughness exponent for interfaces with long-range spatial correlations. While the study of the KPZ equation~\cite{Halpin_Healy_2015} has been fruitful, both the theory and phenomenology of more complex interfaces remain largely unknown. \\

\begin{figure}[htbp!]
    \centering
    \includegraphics[width=1\linewidth]{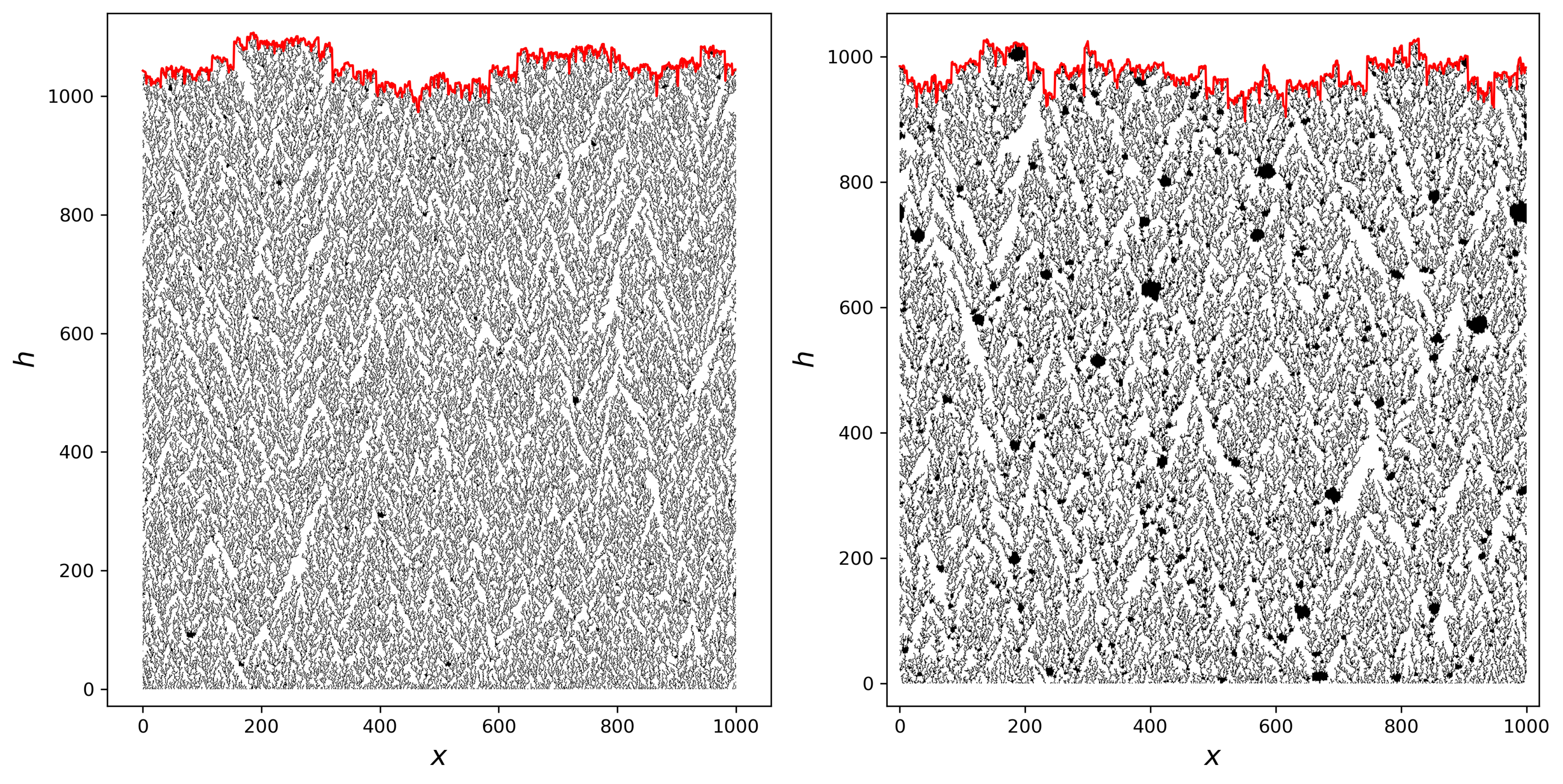}
    \caption{Snapshots of the surface for $\tau=3.5$ (left) and $\tau=2.5$ (right). The system size is $L=1000$, and the number of deposited clusters is chosen such that the average height reaches $h \approx 1000$ in both cases. The red line indicates the final interface.}
    \label{fig:blobs2}
\end{figure}

\section{Model}

The model starts from a flat interface \(h(x,0)=0\) of lateral size \(L\) with periodic boundary conditions. Growth proceeds via the sequential deposition of rigid Eden clusters~\cite{Eden:1961}. At each step, a deposition site is chosen at random, and the blob is lowered vertically until its bottom first contacts the existing surface, either directly or at a neighboring site. Upon contact, the blob adheres irreversibly, without any surface relaxation, as shown in Fig.6 in the Appendix. For unit-size blobs (\(s=1\)), this reduces to the standard next-to-nearest-neighbor~\cite{Barabasi_Stanley_1995} model. Blob sizes (i.e. how many elementary particles they contain) are drawn from a power-law distribution  
\begin{equation} \label{eq:powerlaw}
    P(s) \sim s^{-\tau}, \quad 1 \le s \le L,
\end{equation}
so that for \(\tau \le 2\) there is no characteristic size and the mean size diverges, whereas for \(2 < \tau \le 3\) a typical size emerges but large fluctuations persist, with a divergent second moment. Time is defined as $t \equiv N/L$, representing the average number of blobs deposited per unit length, where $N$ is the total number of deposited blobs and $L$ is the system size. Typical interface roughening is shown in Fig.~\ref{fig:blobs2}, in the cases of~$\tau=3.5$ and~$\tau=2.5$. The variation in frequency of apparition and magnitude of the large blobs can be appreciated.

\section{Exponents}

The statistical properties of growing interfaces are commonly characterized through their roughness,
\begin{equation}
    W^2(L,t) = \frac{1}{L} \sum_{i=1}^L \left[ h_i(t) - \overline{h}(t) \right]^2 ,
\end{equation}
which measures the amplitude of height fluctuations around the mean interface position. For finite systems, the roughness typically grows in time and eventually saturates. In the saturated regime, its dependence on system size defines the roughness exponent~$\alpha$:
\begin{equation}
    \langle W(L,t \to \infty) \rangle\sim L^{\alpha}\,.
\end{equation}

\begin{figure}[htbp!]
    \centering
    \includegraphics[width=1.0\linewidth]{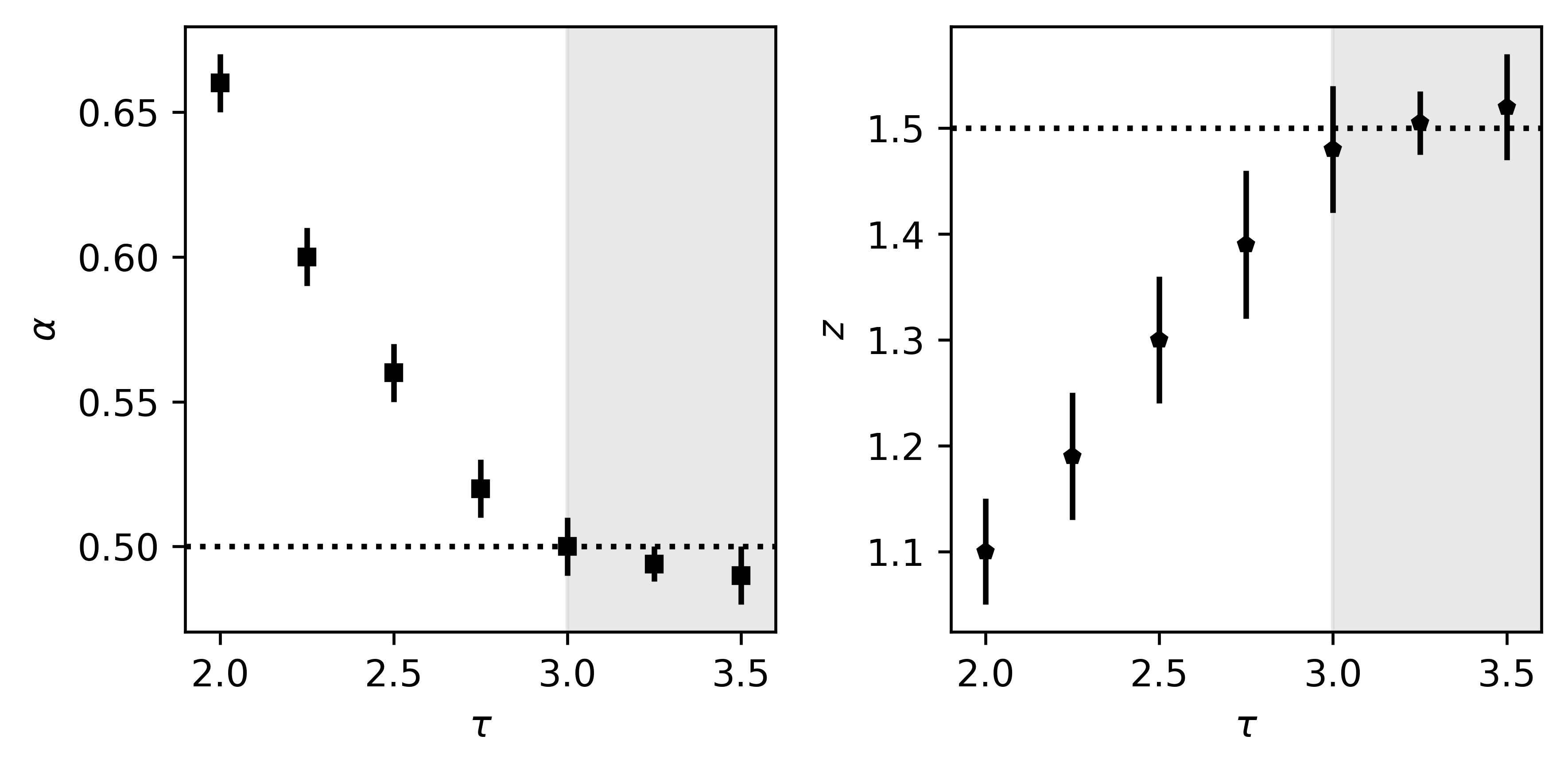}
    \caption{\textbf{Measured roughness and dynamic exponents $\alpha$ and $z$ as a function of $\tau$.}
For $\tau<3$, the critical exponents depend continuously on $\tau$, while for $\tau\geq3$ (highlighted in grey) the KPZ universality class is recovered. Dotted lines indicate the KPZ universal exponents. The determination of the exponents and the associated error bars are discussed in the SM.}
    \label{fig:exponents}
\end{figure}

The characteristic crossover time to saturation scales as $t_{\times} \sim L^{z}$, introducing the dynamic exponent~$z$. Many interfaces also have an early-time roughness scaling algebraically as~$\langle W(t) \rangle \sim t^\beta$, in which case the growth exponent~$\beta=\alpha/z$ follows from the dynamic scaling ansatz. Together, these exponents provide a first level of classification of interface growth processes into universality classes. 

We systematically measured the roughness and dynamic exponents over the range $2 \leq \tau \leq 3.5$. For measuring~$\alpha$, we computed the average long-time limit of the roughness~$W_\text{sat}(L) = \lim_{t \to \infty} \langle W(L,t) \rangle$ and extracted the exponent via the fit~$W_\text{sat}(L) \sim L^{\alpha}$. Instead, the dynamic exponent~$z$ was computed by estimating the saturation time~$t_\times$ for the average roughness to reach a value~$(1-p) W_\text{sat}$ ($p \ll 1$) and fitting~$t_\times \sim L^z$ (more details are given in the Appendix A). The resulting exponent values are summarized in Fig.~\ref{fig:exponents}. In the interval $2 < \tau \leq 3$, where the mean cluster size is finite but its variance diverges, we observe a continuous evolution of the scaling behavior. Specifically, the roughness exponent decreases smoothly from $\alpha(\tau=2)\approx0.66$ to $\alpha(\tau=3)\approx0.5$, following a convex trend, while the dynamic exponent increases from $z(\tau=2)\approx1.1$ to $z(\tau=3) \approx 1.5$.

For $\tau \geq 3$, where the second moment of the cluster-size distribution becomes finite, both exponents converge to their KPZ values. This behavior indicates that the KPZ universality class is recovered whenever fluctuations in cluster size are finite, and that only sufficiently broad, heavy-tailed size distributions are able to drive the system away from KPZ scaling. Consistently, the roughening dynamics of the random Tetris model (see Fig. 10 in the Appendix), which involves deposition of fixed-size random shapes, also falls within the KPZ universality class, providing further support for this interpretation.

In contrast, for $\tau < 3$, the Galilean-invariance relation $\alpha + z = 2$ is clearly violated (see Fig.~8 in the Appendix), signaling a departure from KPZ-type growth. In this respect, earlier results obtained for KPZ equations driven by power-law noise~\cite{zhang1990, zhang1990_2, lam1992, krug1991} do not directly extend to kinetic roughening driven by laterally extended objects. In particular, adapting Krug’s scaling argument~\cite{krug1991} to the present system leads to predictions that are inconsistent with our numerical findings.

\begin{figure}[htbp!]
    \centering
    \includegraphics[width=0.8\linewidth]{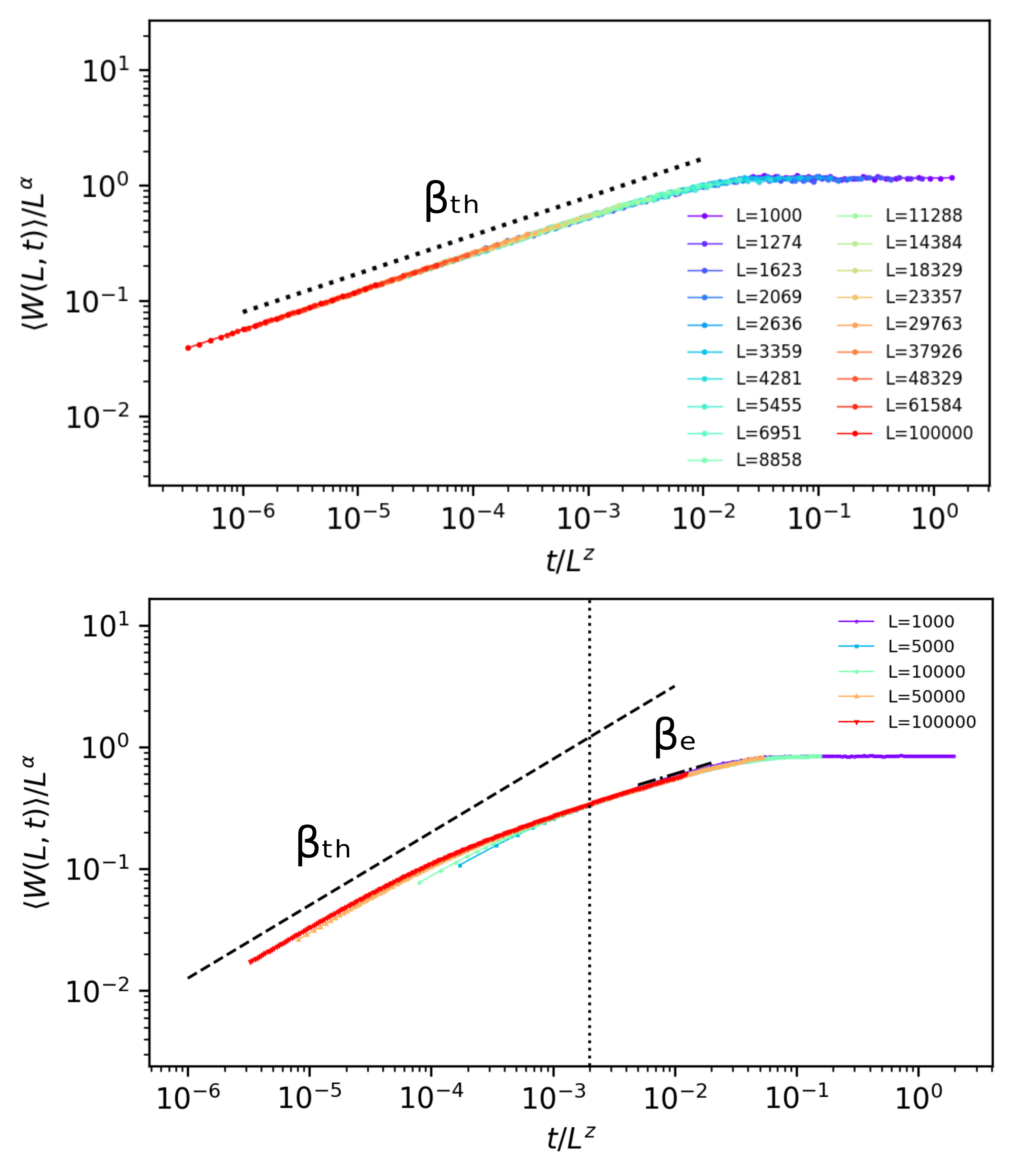}
    \caption{{\bf Dynamic scaling breakdown.} (Top) Family-Vicsek collapse for~$\tau =3$. Dotted line : predicted growth exponent~$\beta_{\text{th}}=1/3$. (Bottom) Failure of the dynamic scaling ansatz. For~$\tau =2$, the FV ansatz does not hold : the curves do not collapse on the whole time range on a master curve upon normalization, and there are strong, system-size dependent corrections to the roughness which effectively does not scale as~$t^{\beta}$ (here~$\beta=0.6$). Close to the saturation point, the curves collapse and can be described by a decreasing effective exponent~$\beta_\text{e}(t)$. The dash-dotted line shows a local tangent to the curve with~$\beta_\text{e} = 0.3$. The vertical dotted line represents the approximate (rescaled) time at which curves collapse again. The dashed line represents the scaling with the exponent $\beta_{th}$ predicted by the Family-Vicsek ansatz. }
    \label{fig:collbreakdown}
\end{figure}

Moreover, simulations of random deposition of rods with broadly distributed lengths characterized by an exponent $\mu$ (see Fig.~9 in the Appendix), yield roughness exponents that clearly deviate from the prediction $\alpha(\mu) = (2+d)/(1+\mu)$. This confirms the numerical observations of \cite{buldyrev1991} and invalidates Krug’s prediction in $1+1$ dimensions. 
It is worth noting that the thresholds for the recovery of the KPZ regime, namely \(\mu=4\) for rods and \(\tau=3\) for blobs in our simulations, are consistent with the formal correspondence obtained by identifying the rod length with the linear size of a compact blob, \(\ell \sim \sqrt{s}\) (see Appendix~B for details).

Altogether, our results show that KPZ universality in the random deposition of clusters remains robust as long as the variance of the cluster-size distribution is finite. Once these fluctuations diverge, both the roughness and dynamic exponents vary continuously with $\tau$, at least down to $\tau=2$.

\section{Effective growth exponent} 

\begin{figure}
    \centering
    \includegraphics[width=0.8\linewidth]{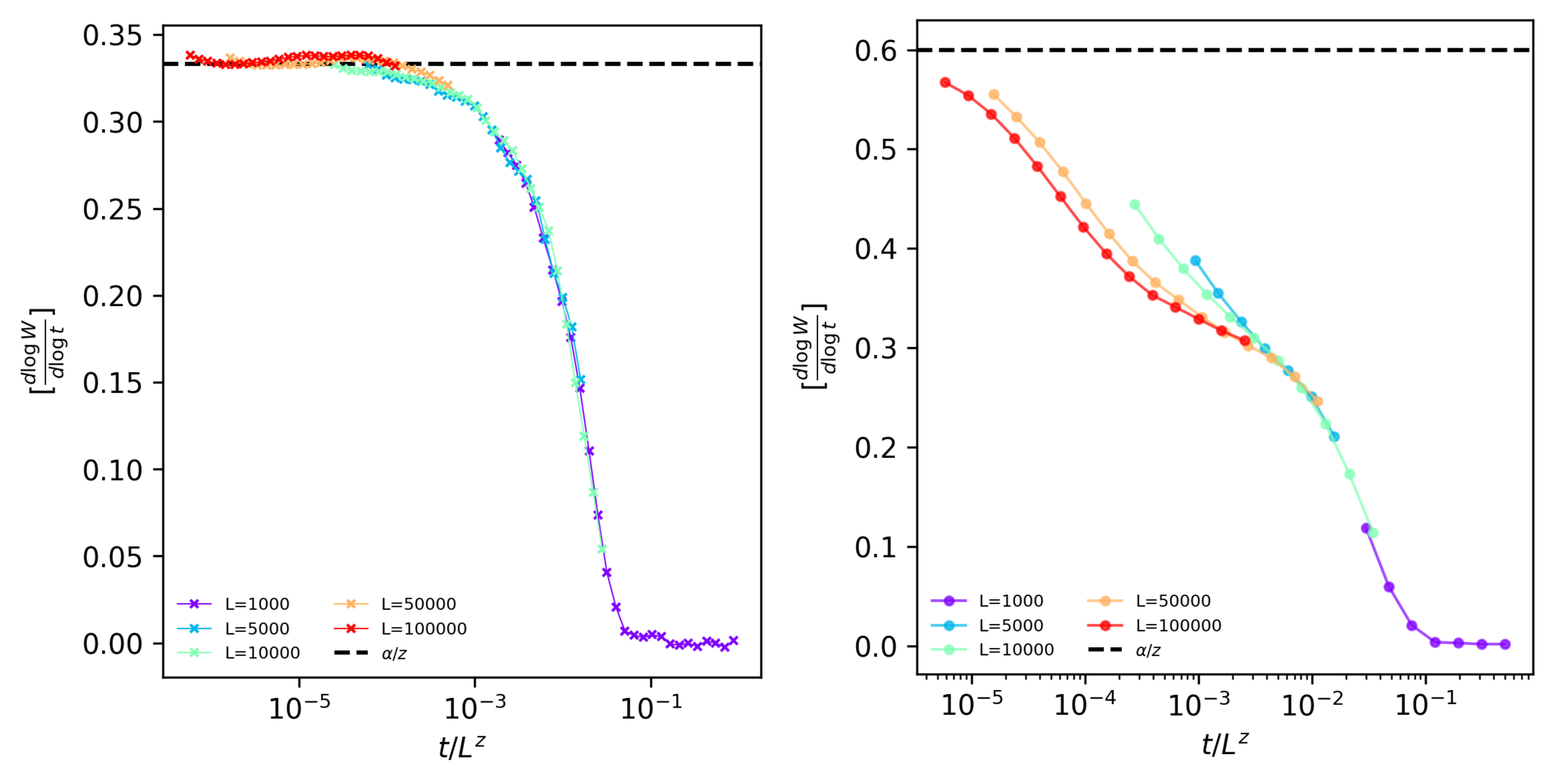}
    \caption{{\bf Effective growth exponent $\beta_e$.} The effective growth exponent~$\beta_e=\mathrm{d} \log W(L,t)/\mathrm{d} \log t$ is shown as a function of the rescaled time~$t/L^z$, for system sizes varying from~$L=10^3$ to~$10^5$. The exponent predicted by the Family-Vicsek ansatz~$\alpha/z$ is displayed as a dashed line. (Left) In the case~$\tau=3$, the effective slopes are independent of the system size and hence collapse on a master curve which is constant at small~$t$, and null at large times. (Right) For~$\tau=2$, (right) the corrections display a complex, system size-dependent shape : firstly, it decays from~$\alpha/z$ slowly until reaching a point~$\beta_c \approx 0.3$. Beyond this point, the curves collapse (upon appropriate time-rescaling by~$L^z$) and there is a crossover to the saturation regime.}
    \label{fig:rough_derivative}
\end{figure}

A common way to assess scale-invariance of interfaces is through the data collapse of the rescaled width~$W(L,t)/L^\alpha$ as a function of the rescaled time~$t/L^z$~\cite{family_1985}. Here, the result of the collapse is shown in Fig.~\ref{fig:collbreakdown} for~$\tau=3$ and~$\tau=2$. In the case $\tau=3$, both the early-time scaling regime $W \sim t^{\beta}$ and the saturation regime are recovered, as expected, as the roughness measured for different system sizes collapses onto a single master curve.
In contrast, for $\tau=2$, the picture changes dramatically. First, the measured roughness exhibits a strong deviation from the expected scaling $W \sim t^{\beta}$: the growth displays pronounced curvature, indicating substantial corrections to the power-law behaviour. Second, these corrections appear not to collapse upon rescaling, as evidenced by the early-time
separation of the curves, and also illustrated in
Fig.~7 (Appendix). This behaviour can be further clarified by considering the effective growth exponent, defined as
\begin{align}
\beta_e(t) = \frac{\mathrm{d}\,\log W}{\mathrm{d}\,\log t},
\end{align}
and shown in Fig.~\ref{fig:rough_derivative} for $\tau=3$ and $\tau=2$. The effective exponent immediately before saturation is approximately $0.3$, markedly smaller than the value $\alpha/z \simeq 0.6$ expected from standard scaling.

In the case $\tau=3$ (left panel of Fig.~\ref{fig:rough_derivative}), the effective slope is $\beta_e = 1/3$ before an abrupt crossover to $\beta_e = 0$, corresponding to the saturation regime. In this case, the growth exponent $\beta = \beta_e$ is well defined, and a clear plateau is observed at long times. In contrast, for $\tau<3$ (and in particular for $\tau=2$, shown here), the effective slope exhibits a slow, system-size-dependent decay, in clear contrast with the prediction of the Family--Vicsek ansatz. This decay persists down to $\beta_e \approx 0.3$, where the effective slopes appear to collapse again, before an abrupt transition to $\beta_e = 0$, signaling the onset of a plateau.
It thus appears that the Family-Vicsek dynamic scaling ansatz fails for the interfaces produced by the random deposition of scale-free clusters. 

\section{Discussion: Competition between two dynamic length scales}

In a very broad class of growth models (EW, KPZ, Mullins-Herring,  etc.), the dynamics is governed by a single growing correlation length $\xi(t)\sim t^{1/z}$ and how it compares to the system size $L$. Assuming that the interface is self-affine, i.e. invariant under
the transformation $x\to b x$, $t\to b^{z} t$, $h\to b^{\alpha} h$, the roughness must scale as 
\begin{align}
W(L,t)=L^{\alpha}\, f\!\left(\frac{L}{\xi(t)}\right),
\label{eq:FV}
\end{align}
which corresponds to the celebrated Family--Vicsek (FV) dynamic scaling ansatz \cite{family_1985}. Equation~\eqref{eq:FV} reflects the fact that, in these systems, only a single relevant length scale grows dynamically, such that the dimensionless ratio $W/L^{\alpha}$ depends solely on the dimensionless quantity $L/\xi(t)$. This situation is recovered here in the case $\tau>3$.

In the growth model considered here for $\tau<3$, deposited clusters follow a broad power-law size distribution. As a consequence, the interface is no longer governed by a single dynamical length scale. In addition to the usual correlation length~$\xi(t)$, the rare but very large absorbed clusters introduce a second, independent, time-dependent scale~$\zeta(t)$, which controls the magnitude and frequency of the extreme height increments associated with the largest blobs. 
Because the dynamics involve an additional extreme-value scale, the interface width is not expected to obey the standard Family--Vicsek (FV) form. According to Buckingham's dimensional analysis theorem~\cite{buckingham1914}, if the roughness is controlled by the system size $L$, the dynamical correlation length $\xi(t)$, and the extreme-value scale $\zeta(t)$, the most general scaling form can be written as
\begin{equation}
    W(L,t)
    =
    L^{\alpha}
    F\left(
        \frac{L}{\xi(t)},
        \frac{\xi(t)}{\zeta(t)}
    \right),
    \label{eq:two_scale_width_scaling}
\end{equation}
where $F$ is a two-variable scaling function. In contrast with the standard
FV ansatz, the dependence on the two dimensionless ratios $L/\xi(t)$ and
$\xi(t)/\zeta(t)$ cannot, in general, be reduced to a single scaling variable (note that testing this scaling form numerically is not straightforward and remains an interesting challenge for future work). The FV collapse is therefore expected to fail when $\zeta(t)$ remains relevant.
This failure appears as strong corrections to scaling in $W(t)$, together with
a scale-dependent effective growth exponent $\beta_e$.

The loss of universality is governed by the interplay between $\xi(t)$ and $\zeta(t)$. When $\xi(t)\gg\zeta(t)$, the size of the absorbed clusters becomes irrelevant and the interface behaves as if depositions were microscopic, so that the standard FV scaling is recovered. When the cluster-size distribution has a finite variance ($\tau>3$), the scale~$\zeta(t)$ becomes negligible compared to~$\xi(t)$ at early times, and the FV regime is rapidly established. In contrast, for heavy-tailed distributions with $\tau<3$, the scale~$\zeta(t)$ grows sufficiently fast to remain comparable to~$\xi(t)$ over an extended temporal range, thereby breaking the FV scaling and modifying the effective roughening exponents.

\begin{figure}
\centering
\includegraphics[width=0.9\linewidth]{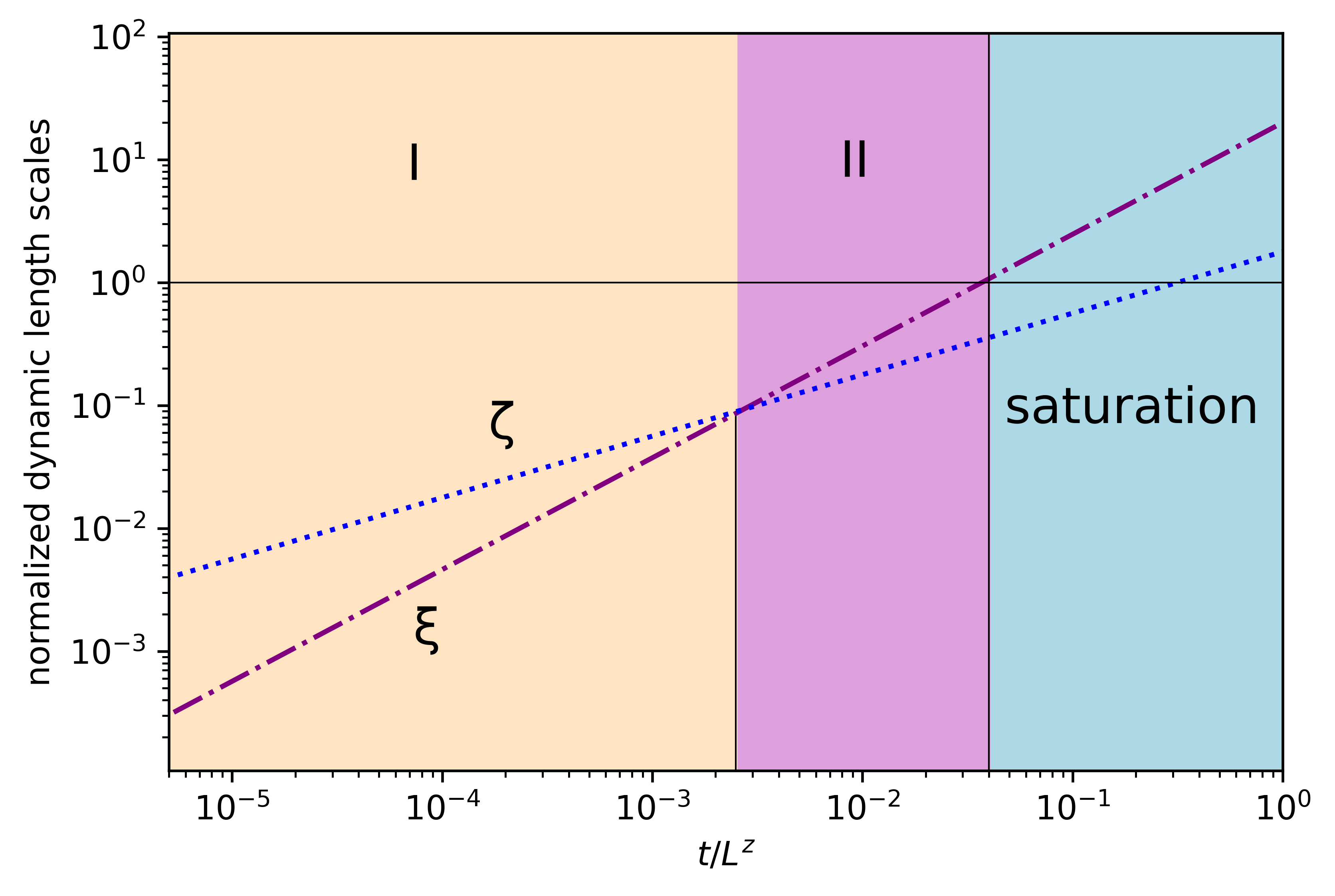}
\caption{{\bf Sketch of the crossover mechanism.}
For illustration we take $\tau=2$.
The correlation length $\xi \sim t^{1/z}$ is shown as a purple dash--dotted line,
using a prefactor inferred from Fig.~\ref{fig:collbreakdown}.
The additional scale $\zeta \sim (L t)^{1/[2(\tau - 1)]}$ (here for $L=10^5$)
is shown as the blue dotted line.
Changing $L$ shifts this curve vertically while preserving its slope.
Axes are rescaled so as to factor out trivial system-size dependence
and to display the saturation region where $\xi/L\sim1$.}
\label{fig:crossover}
\end{figure}

This crossover mechanism is illustrated in Fig.~\ref{fig:crossover}. For $\tau<3$, the line associated with the scale~$\zeta(t)$ has a slope controlled by the tail exponent, and its vertical position depends on the system size~$L$. Changing~$L$ shifts the curve vertically without affecting its slope, and thus alters the intersection point with the correlation-length curve. 

The microscopic origin of the new scale~$\zeta(t)$ follows from the fact that the height at a fixed column is approximately (ignoring porosity) the sum of the vertical increments created by each arriving blob,
\begin{equation}
h(t) \approx \sum_{i=1}^{N(t)} \ell_i ,
\end{equation}
where $\ell_i\sim s_i^{1/2}$ is the vertical size associated with the blob of size~$s_i$, and $N(t)\simeq Lt$ is the number of deposition events. When $\tau<3$, the variance of the cluster-size distribution diverges, and the sum is therefore controlled not by typical events but by its maximal term. By contrast, for $\tau>3$ the variance of $P(s)$ is finite and the central limit theorem applies: no single blob dominates the sum, the scale $\zeta(t)$ becomes negligible, and conventional Family--Vicsek scaling is recovered.

More precisely, for $\tau<3$, extreme-value statistics imply that after $N$ draws
\begin{align}
&s_{\max} \sim N^{1/(\tau-1)}, \\
&\zeta(t) \sim \ell_{\max} \sim s_{\max}^{1/2} \sim N^{1/[2(\tau-1)]} \sim (L t)^{1/[2(\tau-1)]}.
\end{align}
Here \(s_{\max}\) is the largest blob area
or mass sampled up to time \(t\), while \(\ell_{\max}\) is its corresponding
linear size. Thus \(\zeta(t)\) is an extreme-value scale, not a conventional
correlation length. Since the blobs are compact, this linear size characterizes
both their lateral extent and their vertical height scale.

In the regime~$\tau<3$, the column height is controlled by the single largest blob that has appeared up to time~$t$, in close analogy with the generalized central limit theorem~\cite{gnedenko1954limit}. As a consequence, the roughness inherits this dominance of extreme events, generating an additional dynamical length scale that competes with~$\xi(t)$ and prevents collapse onto a Family--Vicsek scaling form.

The roughness dynamics is therefore controlled by the competition between the usual correlation length~$\xi(t)\sim t^{1/z}$ and the extremal scale~$\zeta(t)$. When $\xi\ll\zeta$, fluctuations originating from the arrival of large blobs dominate the growth: the propagation time of correlations is much longer than the spontaneous appearance time of rare events, and the roughness increases through intermittent, large jumps. Once $\xi\gg\zeta$, the situation reverses: the appearance time of such large blobs becomes larger than their effective lifetime, and roughness growth is instead governed by the accumulation of numerous aggregates. 
This crossover mechanism is illustrated in Fig.~\ref{fig:crossover}. For $\tau=2$, three regimes appear. (I) At early times, the roughness growth is dominated by rare, large events for which $\zeta(t)$ exceeds the correlation length. (II) At intermediate times, the increasing propagation range suppresses the effect of newly arriving large blobs. (III) At late times, $\xi(t)$ reaches the system size and saturation occurs.
There exists a critical system size~$L=L_c$ at which the curves $\xi(t)$ and $\zeta(t)$ intersect precisely at the saturation time~$t_\times$. In this case the intermediate regime (II) disappears. For $L<L_c$ the three-regime structure is present; for $L>L_c$, the condition $\zeta(t)>L$ is met before~$t_\times$, which could shift the saturation time to the moment when $\zeta(t)$ reaches~$L$. Such a scenario would break the usual scaling relation $t_\times\sim L^z$ and introduce additional finite-size corrections. Numerically, however, this regime is inaccessible: for $\tau=2$, system sizes of order $10^{15}$ would be required. 

Finally, an additional test of the universality class would be to study the full steady-state width distribution $P(W^2)$ and the maximum relative height distribution $P(h_{\max}-h_{\rm avg})$, which are known analytically for one-dimensional KPZ/EW interfaces \cite{foltin1994width,majumdar2004exact,majumdar2005airy}. Comparing these distributions with the present model would provide a more stringent signature of KPZ behavior, and could also reveal how the system crosses over or departs from this form in the non-KPZ regime for $\tau<3$.

\bibliography{sn-bibliography}

\appendix

\section{Methods}\label{sec11}

\subsection{Eden cluster} \label{subsec:eden}

Clusters deposited on the surface follow the dynamics of an Eden cluster~\cite{Eden:1961, Landau_Binder_2014}. Originally introduced to model the growth of cell colonies, it generates compact, stochastic aggregates whose boundaries roughen over time. Starting from an initial seed, growth proceeds by randomly selecting and occupying one of the empty lattice sites adjacent to the existing cluster. Here, the process is implemented on a regular lattice, in contrast to off-lattice variants commonly used in experimental studies of interface growth, such as those analyzed in~\cite{Takeuchi_2012}.

\subsection{Experimental setup}

\begin{figure}[htbp!]
    \centering
    \includegraphics[width=0.5\linewidth]{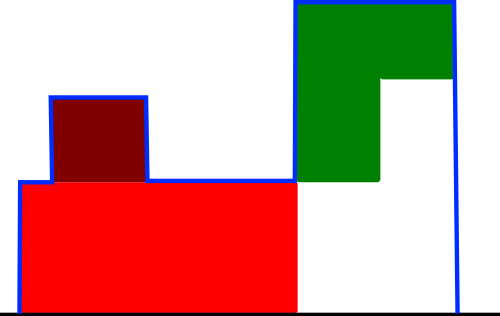}
    \caption{Scheme of the attachment mechanism. The red blob, the first blob, is deposited on the surface~$h=0$ (black line). Then, the green blob, falls with its corner touching the red blob corner. The last blob, the Bordeaux one, is deposited on the red one. There is no relaxation of the blobs. The blue line represents the resulting surface.}
    \label{fig:attachment}
\end{figure}

The surface is initialized at~$h=0$ on a one-dimensional band. Periodic boundary conditions (PBCs) are imposed. Clusters described in the previous section fall on the surface. Their position (i.e, the one of their central seed) is chosen at random, independently from each other. Clusters attach to the surface according to the next-to-nearest-neighbor attachment rule, illustrated in Fig.~\ref{fig:attachment}. The surface coordinates~$h(x)$ is defined as the maximal height reached for a given coordinate~$x$. The system sizes are varied between~$L=10^3$ and~$L=10^5$. For each system size, between~$1000$ and~$5000$ realizations are performed. Depending on the linear extent of the system, the total number of deposited sites ranges from approximately $5 \times 10^{6}$ to $5 \times 10^{9}$, allowing the interface width to be followed over several decades of growth, without making the simulations prohibitively expensive -- for a fixed exponent, simulations (with code written in C) required roughly~$3$ days of computing time on a computing cluster, using around~$20$ CPUs.

\subsection{Kinetic roughening}

The main observable used to characterize the growing interface is the surface roughness, defined as
\begin{equation}
    W^2(L,t) = \frac{1}{L} \sum_{i=1}^L \left( h_i(t) - \overline{h}(t) \right)^2 ,
\end{equation}
where $h_i(t)$ is the height at lateral position $i$ and time $t$, and $\overline{h}(t)$ is the spatially averaged height. The roughness exponent $\alpha$ is extracted from the scaling of the steady--state (saturated) width,
\begin{equation}
    W_{\text{sat}}(L) \sim L^{\alpha} ,
\end{equation}
which provides a straightforward measurement once the interface has reached saturation. 

The dynamic exponent is extracted via scaling the saturation time~$t_\times$ with~$L$. The saturation time~$t_\times$ is estimated through
\begin{align}
    \langle W(L,t_\times) \rangle = (1-p) W_\text{sat} \,.
\end{align}
We varied~$p=0.01,0.05,0.1$ which gives consistent results. Such a method also allows for estimating the confidence intervals around the computed exponent. The measurement of~$z$ is also validated by computing the optimal collapse of the last part of the roughness curves, just before the saturation regime. This allows to compute the exponent but not error bars, at least no straightforwardly. Through this method, the estimated dynamic scaling exponent are consistent with those measured beforehand.

\section{Material}

\subsection{Values of roughness and dynamic exponents}

The measured values of~$\alpha$ and~$z$, shown in Fig.~\ref{fig:exponents}, are displayed in Table~\ref{tab:exponents}.

\begin{table}[h]
\centering
\begin{tabular}{c|c|c|c|c}
\hline
$\tau$ & $\alpha$ & $\alpha$ ($95\%$ CI) & $z$ & $z$ ($95\%$ CI) \\
\hline
2.00 & 0.66 & [0.65, 0.67] & 1.10 & [1.05, 1.15] \\
2.25 & 0.60 & [0.59, 0.61] & 1.19 & [1.13, 1.25] \\
2.50 & 0.56 & [0.55, 0.57] & 1.30 & [1.24, 1.36] \\
2.75 & 0.52 & [0.51, 0.53] & 1.39 & [1.32, 1.46] \\
3.00 & 0.50 & [0.49, 0.51] & 1.48 & [1.42, 1.54] \\
3.25 & 0.49 & [0.48, 0.50] & 1.51 & [1.48, 1.54] \\
3.50 & 0.49 & [0.48, 0.50] & 1.52 & [1.47, 1.57] \\
\hline
\end{tabular}
\label{tab:exponents}
\end{table}

\subsection{Failure of the dynamic scaling collapse}

\begin{figure}[htbp!]
    \centering
    \includegraphics[width=0.7\linewidth]{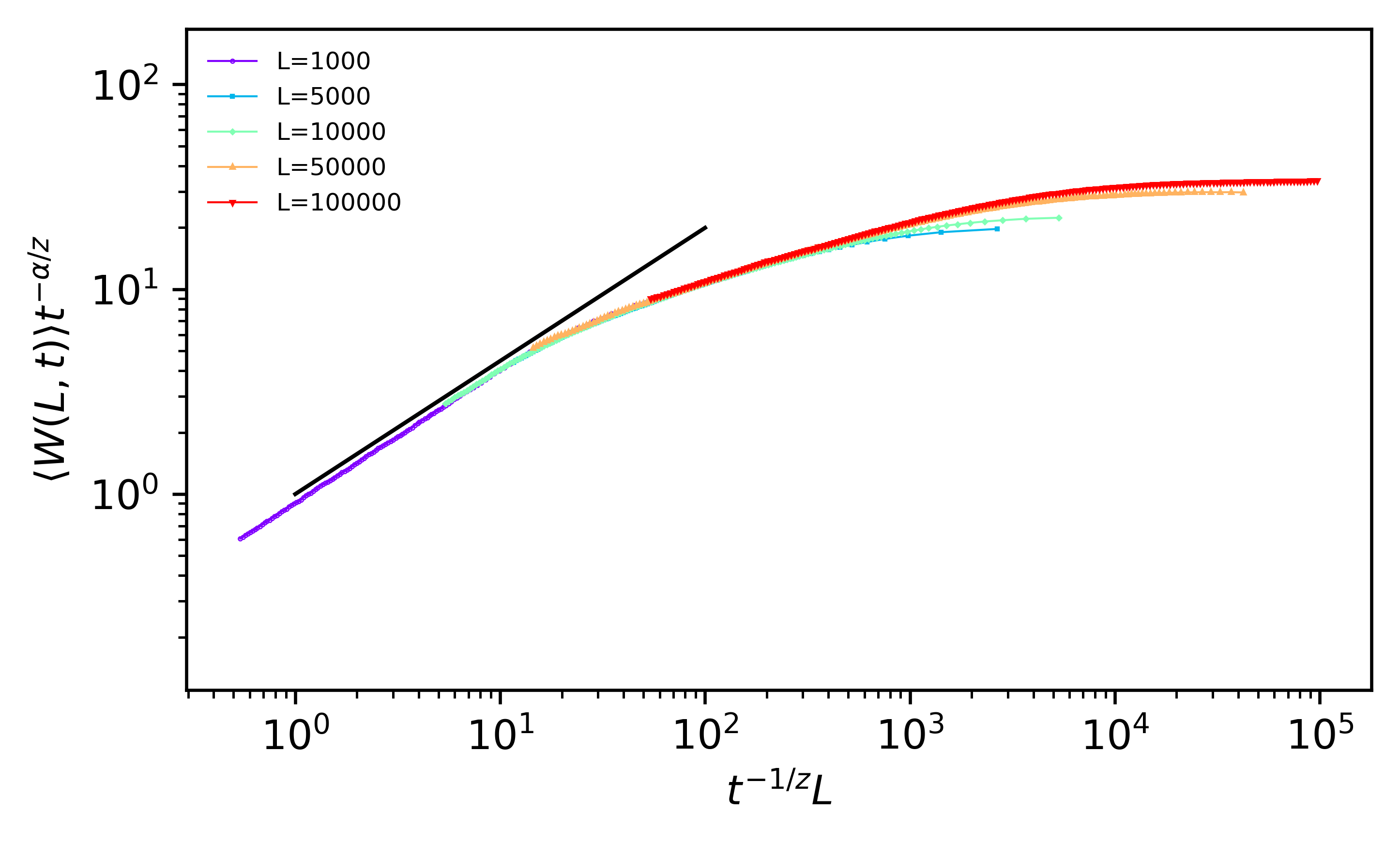}
    \caption{Collapse failure for~$\tau=2$, when using the dynamic scaling form Eq.~\ref{eq:fv2}. The black line is a power-law of exponent~$\alpha$.}
    \label{fig:tau2_sansatzbreaking}
\end{figure}

The Family-Vicsek dynamic scaling ansatz can be written
\begin{equation} \label{eq:fv2}
    W(L,t) = t^{\alpha/z }G(Lt^{-1/z})
\end{equation}
with~$G(u \ll 1) \sim u^{\alpha}$ and~$G(u \gg 1) \sim 1$. In Fig.~\ref{fig:tau2_sansatzbreaking}, we show the result of the collapse when the scaling is written under this form for~$\tau=2$. While the small-$L/t^{1/z}$ behavior is well described by a power-law of exponent~$\alpha$ (the saturation at large times), at large values, we observe a discrepancy, indicating that at small times,~$t^{\alpha/z}$ does not effectively describe the growth of the roughness.

\subsection{Galilean invariance breaking}

\begin{figure}[htbp!]
    \centering
    \includegraphics[width=0.8\linewidth]{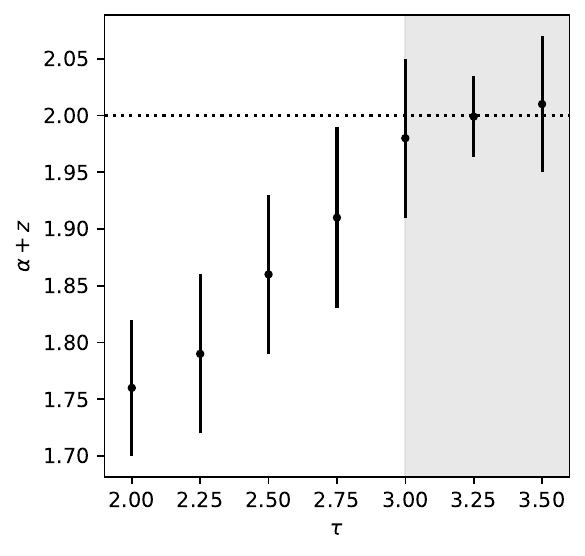}
    \caption{Galilean invariance relation test. Error bars are computed as the sum of the incertitude on the estimations of~$\alpha$ and~$z$. The grey area represents the KPZ part of the phase diagram.}
    \label{fig:galilean}
\end{figure}

The Kardar-Parisi-Zhang equation
\begin{equation}
    \partial_t h = \nu \nabla^2 h + \frac{\lambda}{2}(\nabla h)^2 + \eta
\end{equation}
is determined by symmetry considerations~\cite{kpz, Barabasi_Stanley_1995}. One consequence of these symmetries (when~$\lambda \neq 0$) is the Galilean invariance relation
\begin{equation}
    \alpha+z=2 \,.
\end{equation}
Figure~\ref{fig:galilean} shows the value of $\alpha+z$ as a function of $\tau$ for the model considered here. For $\tau<3$, a strong deviation below $2$ is observed, which varies monotonically with $\tau$. By contrast, for $\tau \geq 3$, the relation $\alpha+z=2$ is satisfied within error bars.

\subsection{Case of rods}

The case of random deposition of rods of length distributed like
\begin{equation}
    P(l) \sim l^{-(\mu+1)}
\end{equation}
was first investigated in a model by Zhang~\cite{zhang1990, zhang1990_2} and then in different cases in~\cite{krug1991,buldyrev1991,lam1992,lam1993}. Krug~\cite{krug1991} conjectured that~$\alpha(\mu)=\frac{2+d}{\mu+1}$, with a critical exponent~$\mu_c=5$ beyond which the roughening falls back to the KPZ universality class. The continuous case was investigated in~\cite{lam1992, lam1993}. 

\begin{figure}[htbp!]
    \centering
    \includegraphics[width=0.7\linewidth]{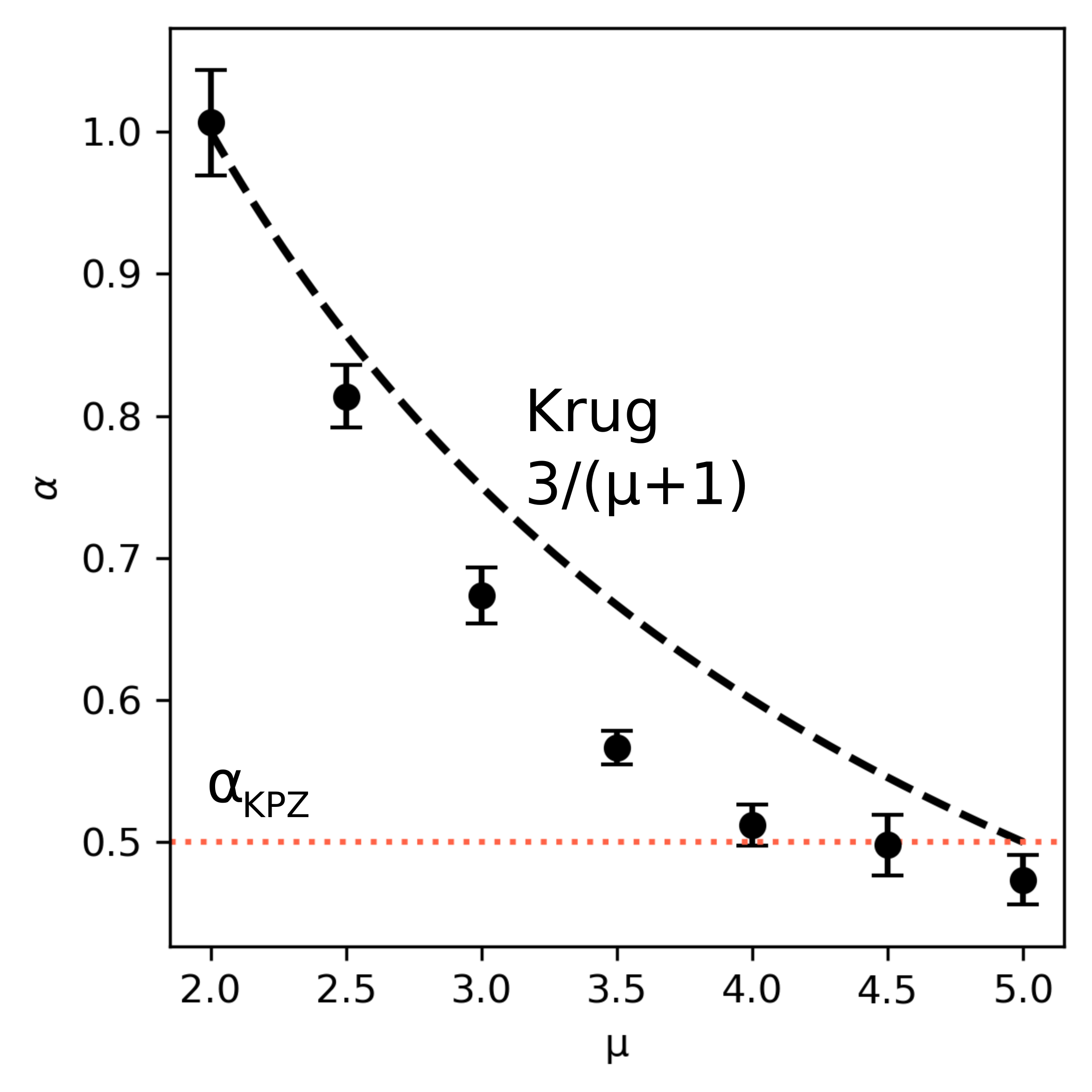}
    \caption{Roughness exponent in the case of rods. Points are measured by scaling the saturated roughness in function of~$L$. The saturated roughness is computed by averaging over~$50$ realizations and the system size varies from~$L=100$ to~$L=10^4$. Error bars are~$95\%$ confidence intervals on the exponent. The dashed line represents Krug's prediction.}
    \label{fig:rods}
\end{figure}

Here, we investigated the model proposed by Buldyrev. We varied the system size between~$L=10^2$ and~$L=10^4$, with~$50$ realizations per configuration, varying~$\mu$ between~$2$ and~$5$. The results are shown in Fig.~\ref{fig:rods}. We find an exponent~$\alpha=1$ at~$\mu=2$, but then the measured exponents decay faster than predicted by Krug, with in particular a saturation at~$\alpha_{\text{KPZ}}=0.5$ already observed at~$\mu=4$, in agreement with results from~\cite{buldyrev1991}.

\paragraph*{A mapping between rods and blobs}

A simple formal correspondence between rod deposition and blob deposition can be
obtained by identifying the rod length with the linear size of a compact blob.
If $s$ denotes the blob size and $\ell$ the rod length, this gives
\begin{equation}
    s \sim \ell^2 .
\end{equation}
Writing \(P(s)\,\mathrm{d}s = P(\ell)\,\mathrm{d}\ell\), and assuming
\(P(s)\sim s^{-\tau}\), we obtain
\begin{equation}
    P(\ell)
    =
    \frac{\mathrm{d}s}{\mathrm{d}\ell} P(s)
    \sim
    \ell \left(\ell^2\right)^{-\tau}
    \sim
    \ell^{-(2\tau-1)} .
\end{equation}
With the definition \(P(\ell)\sim \ell^{-(\mu+1)}\), this gives
\begin{equation}
    \mu = 2(\tau-1).
\end{equation}
Thus, the onset of the KPZ regime at \(\tau=3\) in the blob model corresponds
to
\begin{equation}
    \mu = 2(\tau-1)=4,
\end{equation}
in agreement with the numerical threshold observed for rod deposition.

\subsection{Random tetris}

A useful reference model for kinetic roughening by extended objects is a 
\emph{Tetris-like ballistic deposition} with a simple sticking rule.  We consider a one-dimensional substrate of size $L$ and sequentially drop tetrominoes corresponding to the standard Tetris pieces (I, O, T, L, J, S, Z), chosen with equal probability from a fixed set of orientations. For each deposition event, one tetromino is selected and assigned a random horizontal position such that all of its sites lie within the system. The piece then falls vertically until any of its blocks first touches either the existing interface or the substrate; at that instant it sticks irreversibly, and the local heights of the impacted columns are updated accordingly. No lateral motion or relaxation is allowed after first contact, so that growth is purely ballistic but mediated by extended, anisotropic particles. Time is measured in per-site units $t = N_{\mathrm{tetromino}}/L$, and the interface width $W(L,t)$ is obtained from the standard deviation of the height profile. Despite the geometric complexity of the pieces, the roughness exhibits $W(L,t)\sim t^{\beta}$ at intermediate times with $\beta\simeq 1/3$, and the stationary width scales as $W_{\mathrm{sat}}(L)\sim L^{\alpha}$ with $\alpha\simeq 1/2$, yielding a dynamical exponent $z=\alpha/\beta\approx 3/2$. These exponents are fully consistent with the $1{+}1$ KPZ universality class, showing that KPZ scaling is robust to the deposition of extended tetromino clusters under a purely local sticking rule, provided that fluctuations in the particle size remain small (in particular, have finite variance).

\begin{figure}
    \centering
    \includegraphics[width=0.99\linewidth]{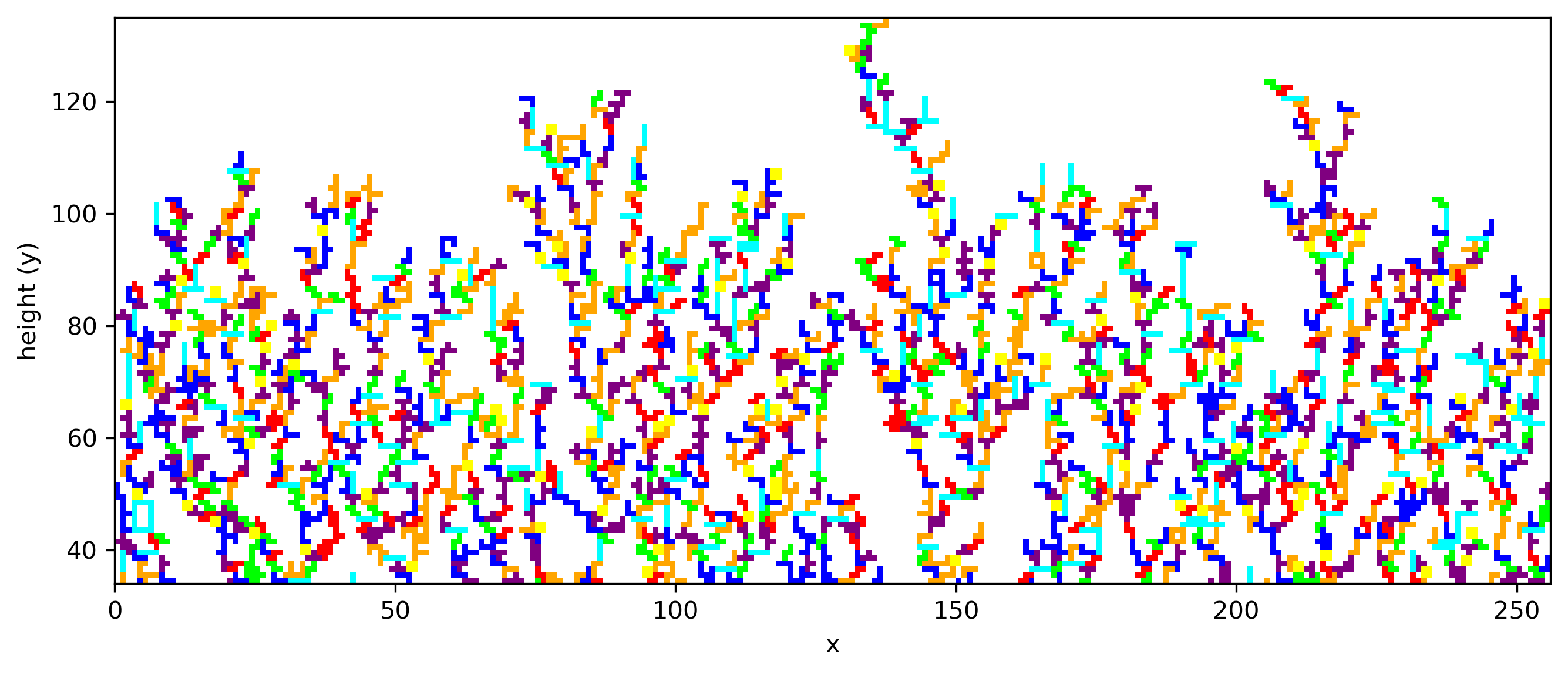}
    \caption{Snapshot of the Tetris-like ballistic deposition interface under the 
    sticking rule. At each step, an extended tetromino (there is a different color for each tetromino) is selected at random, 
    dropped vertically at a random horizontal position, and frozen upon first 
    contact with the existing interface. The complex local morphology generated 
    by these extended particles nevertheless leads to roughening exponents 
    consistent with KPZ scaling in $1{+}1$ dimensions.}
    \label{fig:tetris}
\end{figure}

\end{document}